# Existence of two electronic states in $Sr_4Ru_3O_{10}$ at low temperatures

Zhuan Xu[1], Xiangfan Xu[1], Rafael S. Freitas[2], Zhenyi Long[2], Meng Zhou[3], David Fobes[3], Minghu Fang[1], Peter Schiffer[2], Zhiqing Mao[3], Ying Liu[2,†]

[1] Department of Physics, Zhejiang University, Hangzhou, 310027, P. R. China

[2] Department of Physics and Material Research Institute, the Pennsylvania State University, University Park, PA 16802, U. S. A.

[3] Department of Physics, Tulane University, New Orleans, LA 70118, U. S. A.

[†] Email: liu@phys.psu.edu

We report measurements on in-plane resistivity, thermopower, and magnetization as a function of temperature and magnetic fields on single crystalline $Sr_4Ru_3O_{10}$ grown by the floating zone method. As the temperature was lowered to below around 30 K, the in-plane and *c*-axis resistivities and the thermopower were found to exhibit a step feature accompanied by hysteresis behavior when the in-plane field was swept up and down from below 10 kOe to above 20 kOe. The sharp increase in the thermopower with increasing in-plane magnetic field at low temperatures has not been observed previously in layered transition metal oxides. Comparing with magnetization data, we propose that the step feature marks the transition between the two different electronic states in $Sr_4Ru_3O_{10}$. We propose that the alignment of domains by the in-plane magnetic field is responsible to the emergence of the new electronic states in high applied in-plane magnetic field.

## I. Introduction

Compounds in the Ruddlesden-Popper (R-P) homologous series $Sr_{n+1}Ru_nO_{3n+1}$[1,2], with $n$ = 1 to infinity, exhibit interesting behaviour as the effective dimensionality, determined by the number of perovskite $RuO_6$ layers in a unit cell, $n$, is varied. The $n = 1$ member of the series, $Sr_2RuO_4$, the most two-dimensional (2D) member of the series, is superconducting below 1.5 K[3] featuring an odd-parity, spin-triplet pairing symmetry that was first predicted theoretically[4,5] and confirmed subsequently by many experimental studies[6] including recent phase-sensitive measurements[7]. On the other end of the series, the infinite-layer, three-dimensional member of the series, $SrRuO_3$ ($n$ = infinity) is an itinerant ferromagnet with a $T_c$ of 165 K[8] that may exhibit momentum-space magnetic monopoles[9]. Compounds with intermediate $n$ values, such as $Sr_3Ru_2O_7$ ($n = 2$), are also of fundamental interest because they exhibit interesting phenomena stemming from competition between ferromagnetic (FM) and antiferromagnetic (AFM) instabilities[10,11,12,13] and quantum metamagnetism transition[14,15]. The $n = 3$ compound, $Sr_4Ru_3O_{10}$, has received recent attention because of several interesting observations. So far there seems to be consensus that $Sr_4Ru_3O_{10}$ undergoes a FM transition around $T_c \approx 100$ K based on the measurements carried out on single crystalline samples grown by either the flux[16,17] or the floating zone[18] method. The bulk magnetization was found to start to increase around $T \approx 150$ K[16,17,18], whose origin is yet to be identified, followed by sharp rise around 100 K. The magnetization showed strong anisotropy with respect to its response to the in-pane and $c$-axis fields, reflecting its layered crystalline structure. There is also a strong magnetoelastic coupling, as demonstrated in the Raman spectroscopy measurements[19], and a possible phase separation within a range of the in-plane field[20].

Several basic issues concerning the nature of the FM phase in $Sr_4Ru_3O_{10}$ are not settled however. First, there has been confusion as to whether a second magnetic transition exists below $T_c \approx 100$ K. Magnetic measurements have shown that with the field applied along the $c$ axis, the magnetization as a function of temperature shows a change in slope as the temperature is lowered to below around 50 K[16,17,18]. It was suggested previously that this results from the emergence of an inter-layer AFM phase.

However, there is fundamental difficulty in this interpretation of the data (see below). Second, as the in-plane field is increased, a steep rise in magnetization was observed in the range between 10 – 25 kOe, which was interpreted as a metamagnetic transition possibly associated with spin-reorientation, or evolution of the Fermi surface[21]. However, this conclusion was not supported by neutron scattering measurements[22]. Finally, the saturation field in the in-plane direction is 25 kOe or larger, but only several kOe along the *c* axis, suggesting that the easy axis is along or close to the *c* axis. However, the neutron measurement showed that the microscopic magnetization did not detect any change when the in-plane field was ramped up to 70 kOe, making it difficult to assign the easy axis to the c axis.

The purpose of this paper is to report our detailed measurements on electrical resistivity, thermopower, and magnetization of $Sr_4Ru_3O_{10}$ as the function of the in-plane magnetic field and show that there exist two distinct electronic states at the lowest temperatures as the in-plane magnetic field is varied. We will discuss the physical origin of these two electronic states.

## II. Experimental

Single crystals of $Sr_4Ru_3O_{10}$ were grown by the floating-zone method. The starting materials are $SrCO_3$ (99.99% purity) and $RuO_2$ (99.95% purity). The grown crystals were characterized by X-ray diffraction (XRD) and energy dispersive X-rays (EDX) measurements. The single crystals in this experiment are of a typical size of $2\times1\times0.4\text{mm}^3$. The in-plane residual resistivity $\rho_{ab}$ is about 1.6 μΩ cm at $T$ = 2 K, and the residual resistivity ratio (RRR) is about 160 as compared with 20-30 for crystals prepared by the flux method, a measure of good sample quality. Temperature and field dependence magnetization was measured when $H//ab$ and $H//c$ in a Quantum Design MPMS-5 system. The resistance and thermopower were measured when $H//ab$. Resistance was measured by a standard four-terminal method. Thermopower was measured by a steady-state technique. The electric contacts were made by silver epoxy and the contact resistance was less than 0.5 Ω. The temperature gradient applied to the sample was about 0.5 K/mm and was measured by a pair of differential Type E thermocouples. The effect of magnetic field on the sensitivity of the thermocouples was carefully calibrated by using calibrated Cernox sensors. All the transport property measurements were performed in a Quantum Design PPMS-9 system.

## III. Experimental results

### *A. In-plane electrical resistance and magnetoresistance*

Figure 1 displays the temperature dependence of the in-plane resistivity ($\rho_{ab}$) and its derivative ($d\rho_{ab}/dT$) under zero magnetic field and 50 kOe field along the *ab* plane. The inset shows the in-plane magnetoresistance (MR), measured at 50 kOe, versus temperature for the temperature range between 2 and 30 K. The MR is seen to grow significantly for temperatures below 150 K. The derivative of the in-plane resistivity, $d\rho_{ab}/dT$, shows two peaks under zero field, approximately at 105 K and 35 K. The low-temperature peak vanishes under a 50 kOe in-plane field. In a FM metal, the resistivity as a function of temperature is predicted to feature a change in slope at $T_c$, which should correspond to a jump in $d\rho_{ab}/dT$ but frequently turns into a peak in $d\rho_{ab}/dT$ experimentally. While the high-temperature peak clearly signals an FM transition, the physical origin of the $d\rho_{ab}/dT$ peak at 35 K in zero field is probably not caused by a second FM transition based on the magnetization data (see below), but rather by other change in magnetic structure that contribute significantly to the in-plane electrical transport. The fact that this peak disappears in a 50 kOe in-plane field, however, suggests that the change in the in-plane electrical transport in $Sr_4Ru_3O_{10}$ at this temperature is suppressed by the in-plane field of 50 kOe. Interestingly, there are two sign reversals of in-plane MR around 26 K (rather than 35 K) and 4 K as shown in the inset of Fig.1. An applied magnetic field tends to align spins and suppress spin fluctuation, leading to a negative MR. The fact that the sign reversals in MR and the peak in $d\rho_{ab}/dT$ occur at different temperatures suggest multiple processes are affecting the in-plane electrical transport.

Figure 2 shows the in-plane MR *vs*. $H//ab$ at several temperatures. At the lowest temperatures ($T$ = 2 and 4 K), the in-plane MR increases gradually first when $H$ is lower than 10 kOe, followed by a quick drop when $H$ increasing from 13 to 26 kOe and a level-off as the field increases further. As the temperature increases further, the in-plane MR becomes positive even at the highest fields, with its magnitude decreasing as the field increases. However, the two distinct regimes marked by gradual change in MR are clearly discernible. Interestingly, hysteresis behaviour was found when the system was brought between the two regimes by the up- and down-sweep of the in-plane field. This hysteresis behaviour was observed up to about 30 K, above which the in-plane MR becomes negative in the entire field range. As the temperature increases further, the magnitude of the negative in-plane MR becomes larger. However, the trend is reversed above 50 K. The temperature dependence of in-plane MR and that of $d\rho_{ab}/dT$ described above therefore suggests that fundamental magnetic properties of $Sr_4Ru_3O_{10}$ may still be varying even below $T_c$ = 100 K when the material becomes FM ordered.

*B. In-plane thermopower*

Figure 3 shows thermopower ($S$) *vs*. temperature. Within the whole temperature range, $S$ is positive, increasing smoothly as the temperature was raised, reaching 34 μV/K at room temperature, slightly larger than that of $Sr_2RuO_4$ ($S$ = 29 μV/K)[23]. For conventional semiconductors, positive thermopower would suggest positive charge carriers (holes). However, this conclusion may not hold here given that the thermopower of $Sr_2RuO_4$ is also positive in the temperature range measured (4.2 – 300 K), even though both electrons and holes are known to be present.

The normalized change of thermopower with magnetic field, $\Delta S(H)/S(0)$, exhibits interesting behavior as shown in Figure 4, where $\Delta S(H) \equiv S(H) - S(0)$, and $H$ is applied along the *ab* plane. For $T < 30$ K, $\Delta S(H)/S(0)$ hardly shows any change at low fields, followed by a sharp increase between 8 to 20 kOe, becoming flat as the field increases further. But the crossover between the two regimes in the low and high in-plane field is only slightly larger than that of $\rho_{ab}$. As $T$ is above 50 K, $\Delta S(H)/S(0)$ is seen to increase quickly without a well defined low-field regime. A large hysteresis with the up- and down-sweep of the in-plane field was also observed, again up to about 30 K. No particular feature was found around 50 K, different from the in-plane resistivity behaviors.

*C. Magnetization*

Figure 5a shows the temperature dependence of magnetization ($M$) under a magnetic field of 1 kOe applied along either *c* axis or *ab* plane. The measurements were carried out on crystals that were also grown by the floating zone method, but different pieces from those used in the electrical transport and thermopower measurements. With the magnetic field applied along the *c* axis, $M(T)$ shows a sharp increase at about 100 K, which is identified as the $T_c$ for the FM transition in $Sr_4Ru_3O_{10}$. Below around 65 K, which is slightly larger than the temperature at which the in-plane MR trend changes (50K) as discussed above, however, $M(T)$ displays a change of slope, increasing more quickly than at higher temperatures as $T$ is lowered. In addition, $M(T)$ curves for the field-cooling (FC) and zero-field-cooling (ZFC) is seen to deviate from one another at this temperature, 65 K. For magnetic field applied in an in-plane direction (the field was not aligned with any specific in-plane direction), $M(T)$ shows a pronounced peak around 65 K, decreasing sharply with the decreasing field until it levels off around 30 K. Moreover, different from the case in which the field was aligned along the *c* axis, there is little difference between FC and ZFC curves.

Figure 5b shows the magnetization as a function of in-plane field, *H*//*ab*, at several temperatures as indicated. At the low temperatures, for example, $T$ = 2 K, $M$(*H*//*ab*) is seen to increase roughly linearly at the lowest field, before rising more sharply in the range of 15 – 25 kOe. Above *H*//*ab* ≈ 25 kOe, the magnetization increases much gradually as the field increases further. Again, a hysteresis behavior for the up- and down-sweep of the in-plane field was observed in the crossover between the two regimes. Above 30K, the hysteresis disappears, and the magnetization displays behavior of a typical ferromagnet.

Similar behaviour of $M(T)$ as well as $M(H)$ was observed in previous reports[16,17,18]. Interestingly, the exact shape of the slope change in $M(T)$ and difference between the FC and ZFC behaviours with a *c*-axis field appear to be highly sample-dependent. On the other hand, the behaviour in $M(T)$ appears to be insensitive to details of the sample.

*D. c-axis resistivity under an in-plane field*

The *c*-axis resistivity as a function of in-plane field, $\rho_c$(*H*//*ab*), at various temperatures is shown in Figure 6. Below 30 K, $\rho_c$(*H*//*ab*) first increases in the small field regime, decreases surprisingly sharply near the field ranging between 20 – 25 kOe (depending on temperature), and then varies much more gradually as the in-plane field increases further. Small hysteresis was seen below 30 K as well, as shown in the inset of Fig. 6a for $T$ = 2 K. The small-field regime appears to disappear above 30 K. However, even at 45 K, the drop in *c*-axis resistivity is still very sharp, especially in comparison with similar step feature observed in the in-plane resistivity shown in Fig. 2.

## IV. Discussion

As shown above, even though $M(T)$ with the field applied along the *c* axis showed a steep rise around 65 K, $M(T)$ as a function *H*//*ab* actually starts to decrease below around the same temperature, reaching a value close to zero at the lowest temperatures when the field is applied along the in-plane direction. Similar behaviour in $M(T)$ was observed previously. It was proposed that a FM intraplane ordering without inter-layer coupling emerges below 100 K, and an AFM ordering among the FM layers below the second characteristic temperature, about 50 K[17,19] (65 K in the preparation). The main problem with the above interpretation is that an interlayer AFM ordering will lead to a vanishing $M(T)$ measured with an in-plane field only if the field is aligned along the easy axis. Since the crystal axes in the magnetic measurements, including these reported here, were not aligned with respect to the field, this interlayer AFM ordering scenario is not likely to be true. Indeed, no AFM ordering of any kind was observed in the neutron scattering study.

The magnetization measurements showed that the saturation field along the in-plane direction is 25 kOe or larger, but that along the *c* axis is only several kOe, suggesting that the easy axis is along or close to the *c* axis. However, the neutron measurement showed that the microscopic magnetization did not detect any change when the in-plane field was ramped up to 70 kOe within the experimental resolution, making it difficult to assign the easy axis to the *c* axis direction. Therefore, if the step like feature observed in various measurements as described above is indeed due to a metamagnetic transition tuned by the in-plane field, the field-induced magnetic moment must be small in comparison with the moment responsible for spontaneous ferromagnetic ordering.

Physical insight may be obtained by considering the strong magneto-elastic coupling found in a Raman study carried out previously in $Sr_4Ru_3O_{10}$. In that study, the 380 $cm^{-1}$ $B_{1g}$ phonon frequency which is associated with internal vibrations of the $RuO_6$ octahedra is highly sensitive to the ferromagnetic order. A distinct change in the slope of the $B_{1g}$ phonon frequency, $d\omega/dT$, is observed below $T_c$ of 105 K. When the magnetic field is applied along the c-axis direction at low temperature (much lower than $T_c$), the Ru magnetic moments are easily aligned to the c axis by the field, and the $B_{1g}$ phonon frequency exhibits a frequency increase with field. The increase in the $B_{1g}$ phonon frequency implies that there is an increase in the elongation of the $RuO_6$ octahedra along the c-axis and a contraction of the in-plane Ru-O bonds under the c-axis magnetic field. Meanwhile the $B_{1g}$ phonon frequency exhibits a significant decrease with increasing in-plane field up to about 20 kOe as $T < 50$ K (actually this effect became clearly visible only for $T < 30$ K), indicating that there is a distinct increase in the in-plane Ru-O bonds and a decrease in the elongation of the $RuO_6$ octahedra along the *c* axis. The decrease in the frequency is more abrupt as the field is around 20 kOe at which the step features in the MR and *S* were also observed in the present work. This was previously interpreted as a metamagnetic transition from AFM to FM ordering.

We wish to present in this paper a picture based on domain structure that may explain the results obtained in the present work. In this picture, we envision the existence of two types of domains in $Sr_4Ru_3O_{10}$ - the "soft" and the "hard" domains. While "soft" domains are expected below $T_c$ of 100 K, "hard" domains start to form only below 50 - 65 K, growing in size as the temperature is lowered. Below 30 K, these hard domains are pinned, and can only be aligned by a large in-plane magnetic field. Because of a strong magneto-elastic coupling, the electronic state of the material can be changed when the entire sample is polarized into a single domain in a strongly enough in-plane magnetic field. For fields applied in the *c*-axis direction, the sample can be polarized into a single domain with a field much smaller than that in the in-pane direction.

The presence of two different types of domain structures could originate from the strong anisotropy. At low temperatures, "hard," two-dimensional-shaped domains are needed to cancel the in-plane magnetization. A small in-plane magnetic field (1 kOe in our case) is not enough to align these "hard" domains as $T < 30$ K, therefore a vanishing $M(T)$ is observed due to the cancellation. The magnetization increases nonlinearly with increasing in-plane field and changes dramatically through domain rotation when the field is in the range between 10 kOe and 20 kOe. A well aligned domain structure forms when the field is large enough. The hysteresis observed in $M$ vs. $H \parallel ab$ curves below 30

K is a consequence of the domain rotation by applied magnetic field. The hysteresis in the *ab*- and *c*-axis MR, and magneto-thermopower could have resulted from the hyteresis in $M(H)$ through a magnetoelastic coupling. For $H \parallel c$, a small field is enough to align the magnetic moments along c-axis direction so that the $M(T)$ is larger compared to the case for $H \parallel ab$ and the vanishing behaviour of $M$ is not significant as $T$ close to zero.

The details of the features found in MR and thermopower support the domain picture. The low-field regime disappears when $T$ is above 30 K, indicating that the domains become "soft" at high temperature. The sharp change in the c-axis MR occurs at slightly higher magnetic fields (25 -30 kOe) than for the in-plane resistivity. The extremely sharp drop in the c-axis resistivity below 30 K may indicate that a single domain might form at high magnetic field. Actually the c-axis resistivity drop was observed up to a temperature as high as 70 K, but the low-field shoulder was found to disappear at $T > 30$ K, similar to the case of in–plane resistivity, indicating that the low-field electronic state becomes less well defined above 30 K.

Dramatic changes in heat capacity[24] were also observed below 30 K when the in-plane magnetic field was raised. Recently specific heat measurements performed on the $Sr_4Ru_3O_{10}$ single crystals grown by flux method showed that the specific heat increases gradually in the low field range, jumps up sharply at $H_c = 29$ kOe, and decreases slowly as $H \| ab$ increases further. This result supports the existence of two distinct electronic states in $Sr_4Ru_3O_{10}$ in low and high in-plane magnetic fields. We note, however, the above cited $H_c$ is larger than that found in the present work. Such a difference in $H_c$ could be due to difference in samples because the domain structure is also sensitive to possible defects and strains in the crystals.

The increase of thermopower with an increasing magnetic field, as observed in the present work on $Sr_4Ru_3O_{10}$, is quite remarkable. It has been reported that the thermopower in the cobaltate $Na_xCoO_2$ is strongly suppressed by applied magnetic field because the spin entropy which accounts for the large thermopower in cobaltates is efficiently suppressed by the strong magnetic field[25]. Such an increase in thermopower has never been reported in the R-P materials. Therefore, in addition to providing support that a sufficiently strong in-plane magnetic field leads to the emergence of an electronic state different from that in the zero field in $Sr_4Ru_3O_{10}$, our observation suggests that this new electronic state features enhanced entropy. Since the alignment of domain tends to reduce entropy, the increased entropy may have come from orbital rather than spin degrees of freedom, which is very different from the case of cobaltates. However, exactly why the high magnetic field state increases thermopower in $Sr_4Ru_3O_{10}$ is not yet well understood.

## V. Conclusion

We have measured the temperature and in-plane field dependence of the resistivity, magnetization, and thermopower in $Sr_4Ru_3O_{10}$. It was found that there was a step feature in all these physical properties with increasing in-plane magnetic field below $T < 30$ K, below which significant hysteresis was observed as the in-plane field was swept up and down. We propose here that the step feature marks the transition between the two different electronic states in $Sr_4Ru_3O_{10}$. While metamagnetic transition could account for the existence of the two electronic states, we point out here that the two electronic states could correspond to the single- or multiple-domain states in $Sr_4Ru_3O_{10}$ that features a strong magneto-elastic coupling. When a sufficiently strong in-plane magnetic field aligns the “hard” domains to form a single domain, a corresponding change in the lattice structure takes place because of the magneto-elastic coupling, resulting in a new electronic state. Finally, a sharp increase in the thermopower with increasing in-plane magnetic field at low temperatures, which is highly unusual, was observed.

**Acknowledgements.** The authors would like to thank Drs. W. Bao and L. Balicas, and Prof. M. Salamon for useful discussions. The work is supported at Zhejiang University by the Natural Science Foundation of China under Grant No. 10634030 and 10628408, at Tulane University by Research Corporation and NSF under grant DMR-0645305, and at Penn State by DOE under DE-FG02-04ER46159 and by NSF under grants DMR-0401486, INT 03-40779.

**References:**

[1] S.N. Ruddlesden and P. Popper, Acta Cryst. **10**, 538 (1957).

[2] W. Tian, J. H. Haeni, and D. G. Schlom, E. Hutchinson, B. L. Sheu, M. M. Rosario, P. Schiffer, and Y. Liu, M. A. Zurbuchen, X.Q. Pan, Appl. Phys. Lett. 90, 022507 (2007).

[3] Y. Maeno, H. Hashimoto, K. Yoshida, S. Nishizaki, T. Fujita, J.G. Bednorz, and F. Lichtenberg, Nature (London) **372,** 532 (1994).

[4] T.M. Rice and M. Sigrist, J. Phys.: Condens. Matter **7**, L643 (1995).

[5] G. Baskaran, Physica B 223&224, 490 (1996).

[6] HA. P. MackenzieH and HY. MaenoH, Rev. Mod. Phys. 75, 657 (2003).

[7] K. D. Nelson, Z. Q. Mao, Y. Maeno, and Y. Liu, Science **306**, 1151 (2004).

[8] A. Kanbayasi, J. Phys. Soc. Japan 41, 1876 (1976).

[9] Z. Fang, N. Nagaosa, K.S. Takahashi, A. Asamitsu, R. Mathieu, T. Ogasawara, Hi Yamada, M. Kawasaki, Y. Tokura, and K. Terakura, Science 302, 92 (2003).

[10] Y. Liu, R. Jin, Z. Q. Mao, and K. D. Nelson, M. K. Haas and R. J. Cava, Phys. Rev. B **63**, 174435 (2001).

[11] L. Capogna, E. M. Forgan, S. M. Hayden, A. Wildes, J. A. Duffy, A. P. Mackenzie, R. S. Perry, S. Ikeda, Y. Maeno, and S. P. Brown, Phys. Rev. B 67, 12504 (2003).

[12] K. Kitagawa, K. Ishida, R. S. Perry, T. Tayama, T. Sakakibara, and Y. Maeno, Phys. Rev. Lett. 95, 127001 (2005).

[13] J. Hooper, M. H. Fang, M. Zhou, D. Fobes, N. Dang, Z. Q. Mao, C. M. Feng, Z. A. Xu, M. H. Yu, C. J. O'Connor, G.. J. Xu, N. Andersen, and M. Salamon, cond-mat/0607414, to appear in PRB (Rapid Comm.)

[14] S.A. Grigera, R.S. Perry, A.J. Schofield, M. Chiao, S.R. Julian, G.G. Lonzarich, S.I. Ikeda, Y. Maeno, A.J. Millis, A.P. Mackenzie, Science 294, 329 (2001).

[15] A.J. Millis, A.J. Schofield, G.G. Lonzarich, and S.A. Grigera, Phys. Rev. Lett. 88, 217204 (2002).

[16] M.K. Crawford, R.L. Harlow, W. Marshall, Z. Li, G. Cao, R.L. Lindstrom, Q. Huang, and J.W. Lynn, Phys. Rev. B 65, 214412 (2002).

[17] G. Cao, L. Balicas, W.H. Song, Y.P. Sun, Y. Xin, V.A. Bondarenko, J.W. Brill, and S. Parkin, Phys. Rev. B 68, 174409 (2003).

[18] M. Zhou, J. Hooper, D. Fobes, Z.Q. Mao, V. Golub, C.J. O'Connor, Mat. Res. Bull. 40, 942 (2005).

[19] HR. GuptaH, HM. KimH, HH. BarathH, HS. L. CooperH, and HG.. CaoH, Phys. Rev. Lett. **96**, 067004 (2006).

[20] HZ. Q. MaoH, HM. ZhouH, HJ. HooperH, HV. GolubH, and HC. J. O'ConnorH, Phys. Rev. Lett. 96, 077205 (2006).

[21] Y.J. Jo, L. Balicas, N. Kikugawa, K. Storr, M. Zhou, and Z.Q. Mao, cond-mat/0605504 (2006)

[22] HW. BaoH, HZ. Q. MaoH, HM. ZhouH, HJ. HooperH, HJ. W. LynnH, HR. S. FreitasH, HP. SchifferH, HY. LiuH, HH.Q. YuanH and HM. SalamonH, cond-mat/0607428 (2006).

[23] H. Yoshino, K. Murata, N. Shirakawa, Y. Nishihara, Y. Maeno, T. Fujita, J. Phys. Soc. Jpn. 65, 1548 (1996).

[24] G.. Cao, S. Chikara, J.W. Brill and P. Schlottmann, Phys. Rev. B 75, 024429 (2006)

[25] Yayu Wang, Nyrissa S. Rogado, R. J. Cava, and N. P. Ong, Nature (London) 423, 425 (2003).

## Figure captions

Fig 1. The temperature dependence of the in-plane resistivity and its derivative under zero and 5 T in-plane magnetic field. Inset: magneto-resistance versus temperature for the temperature range from 2 K to 30 K.

Fig 2. Magnetic field dependence of the in-plane resistivity at several temperatures as indicated. The magnetic field was applied along the *ab* plane.

Fig. 3. Thermopower (*S*) *vs*. temperature under zero and 4 T in-plane magnetic field. This piece of the $Sr_4Ru_3O_{10}$ sample is the same one used for resistivity and MR measurements.

Fig.4 Magnetic field dependence of the normalized thermopower at several temperatures as indicated. The magnetic field is applied along the *ab* plane. The thermopower at zero field are 0.283, 0.798, 1.49, 2.66, 4.69, 8.32, 9.57, 11.4, 17.6, 19.4, 26.4, 31.5 μV/K, for $T$ =2, 4, 6, 8, 12, 18, 20, 25, 35, 50, 100, 150 K, respectively.

Fig. 5. The temperature dependence of magnetization under both in- and out–plane magnetic fields (a) and field dependence of magnetization under in-plane magnetic fields (b).

Fig. 6. The field dependence of the normalized *c*-axis resistivity at several temperatures as indicated. The inset in (a) shows the hysteresis at $T$ = 2 K. The hysteresis disappears as $T$ > 25 K. The values of $\rho_c(0)$ are (to be input) for T = 2, 4, 8, 15, 25, 35, 45, 55, 70 K, respectively.

Fig.1:

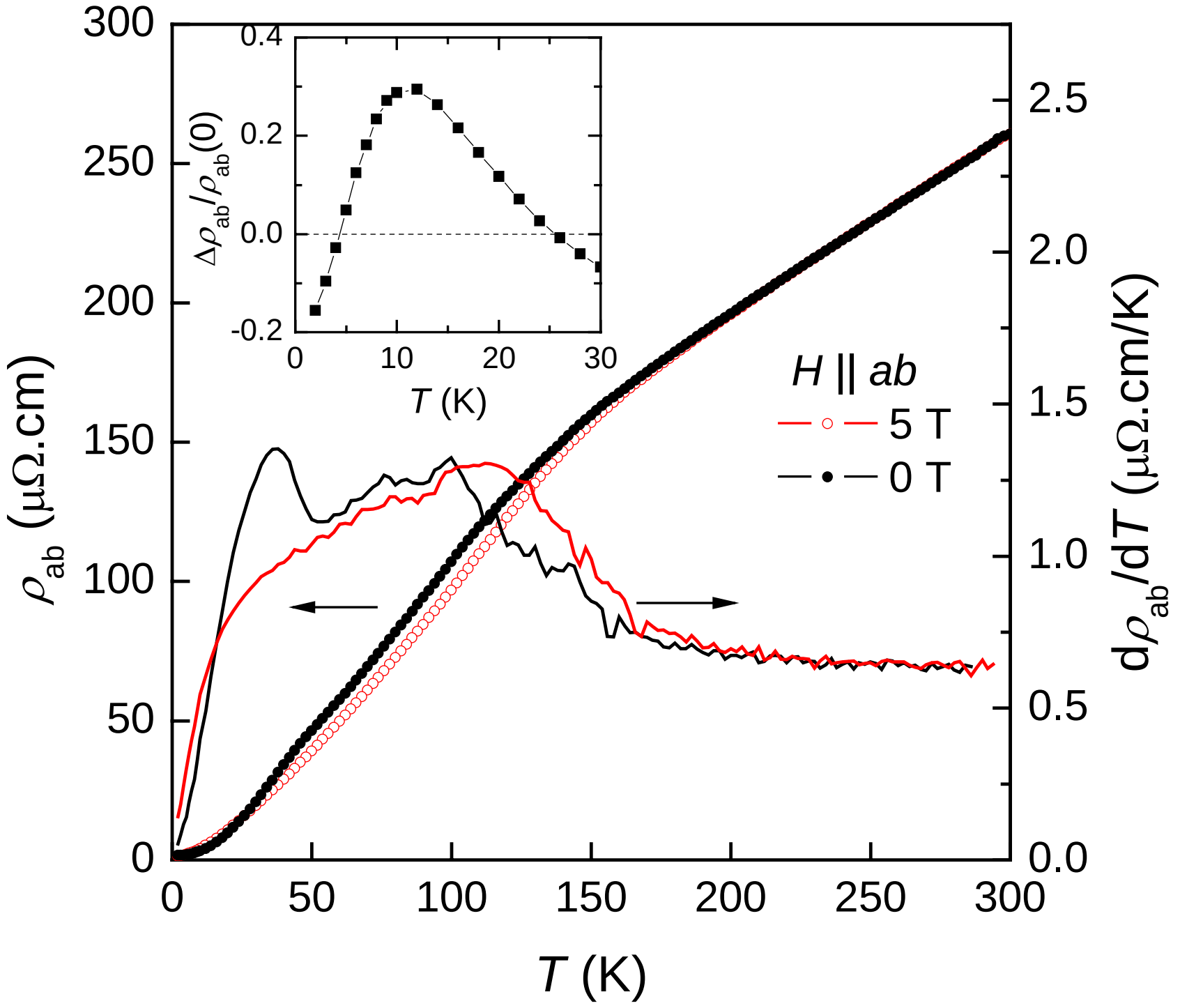

Fig.2

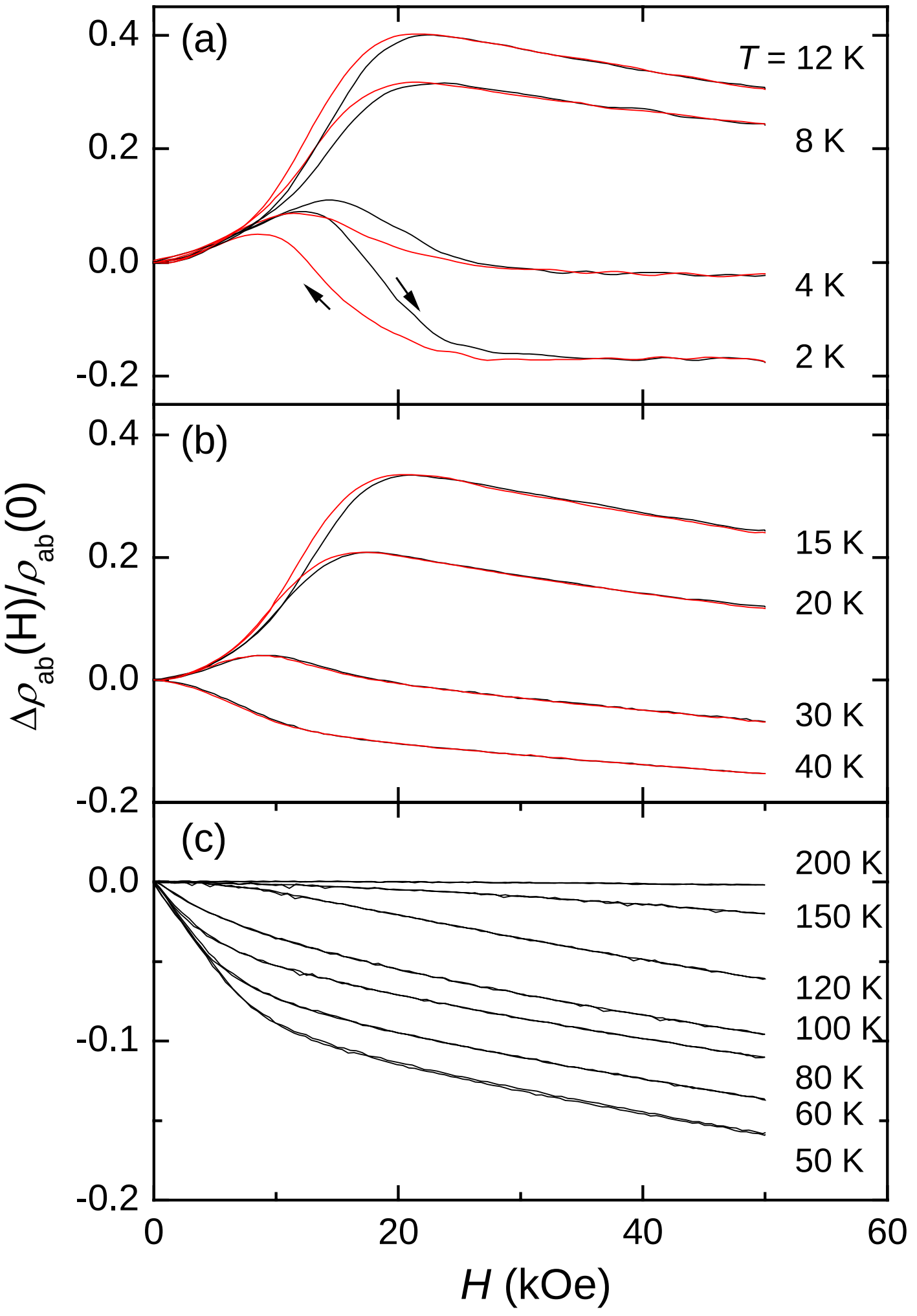

Fig.3

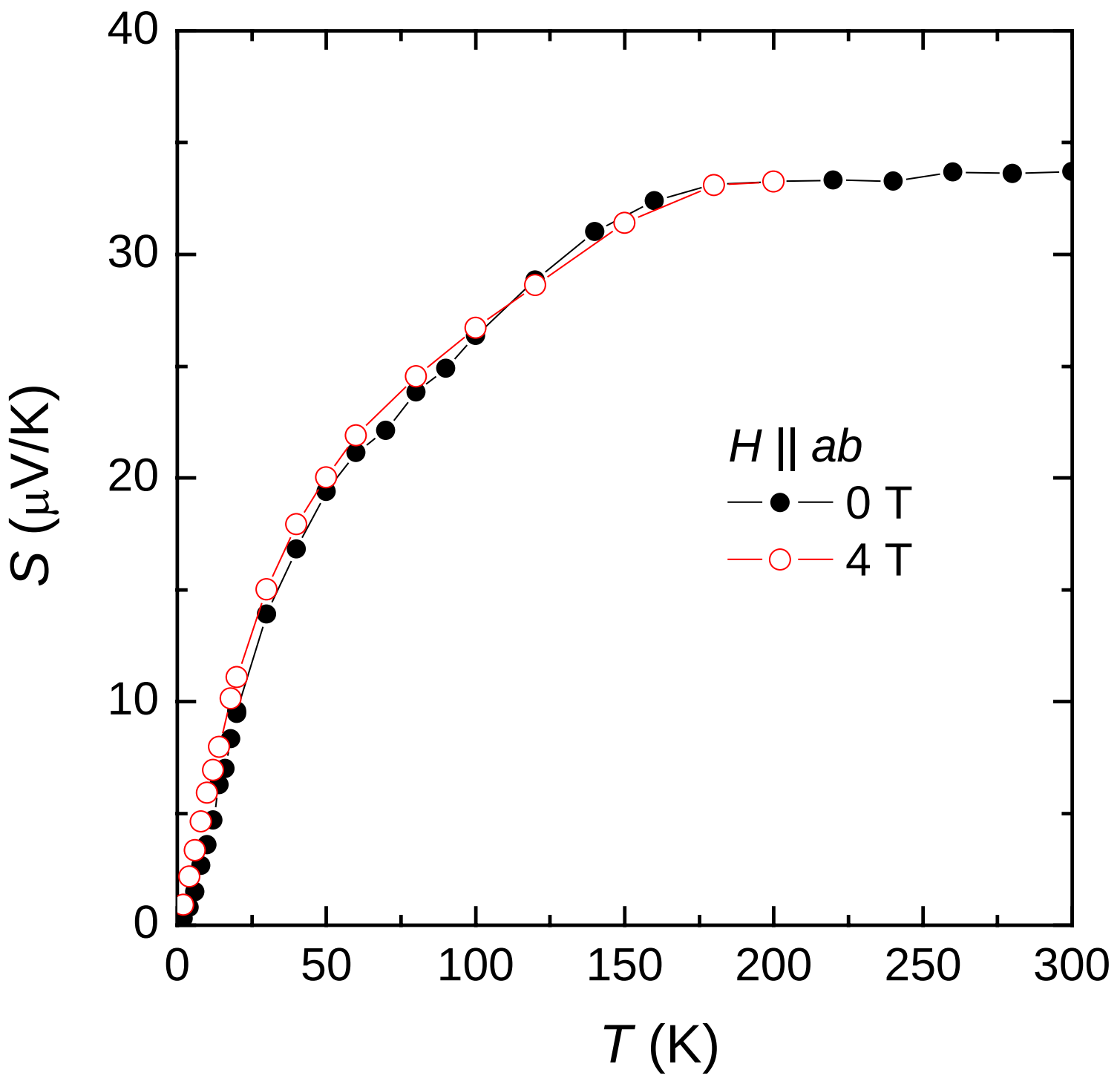

Fig.4

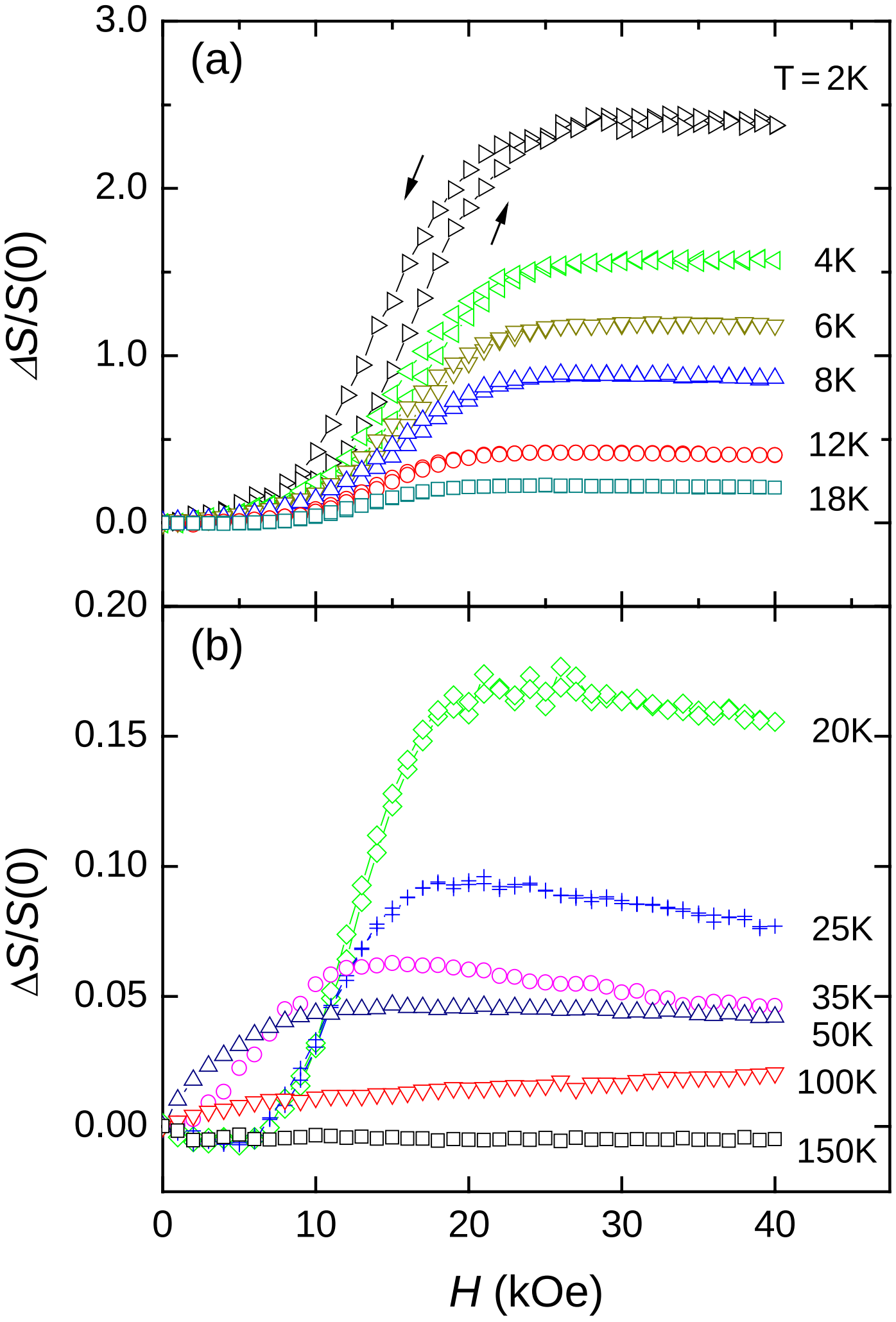

Fig.5

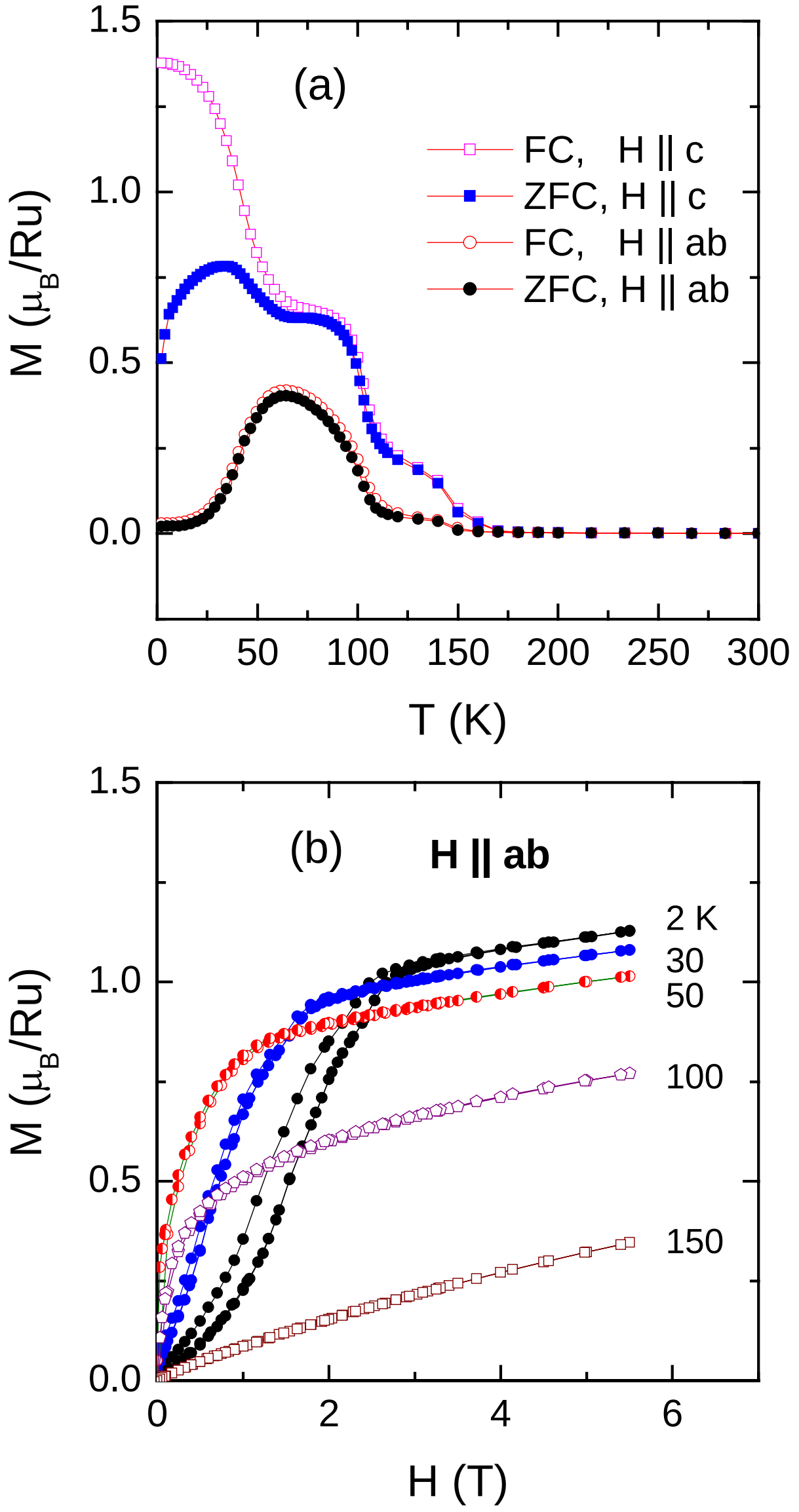

Fig. 6

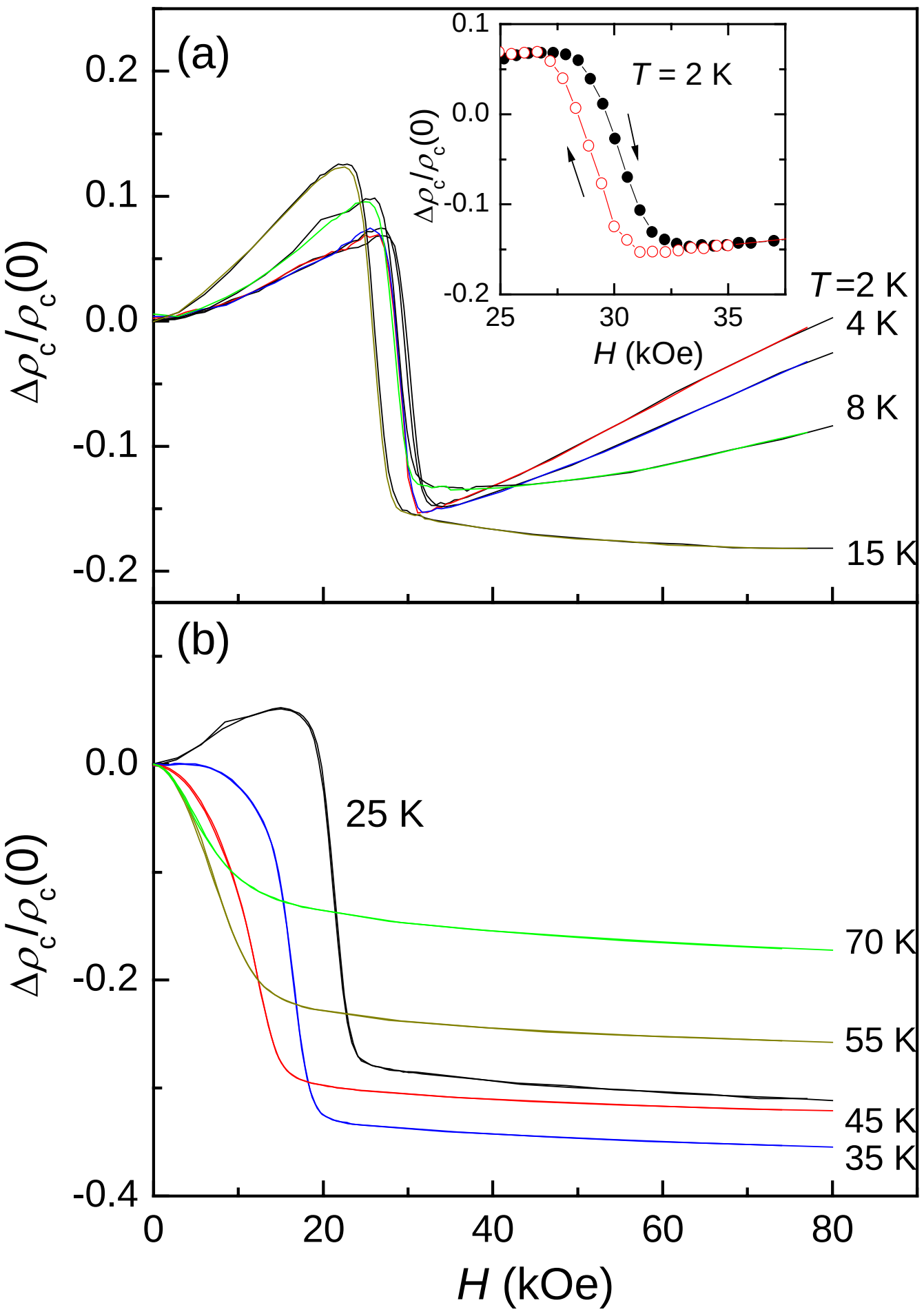